# The Meaning of Group Delay in Barrier Tunneling: A Re-Examination of Superluminal Group Velocities

Herbert G. Winful

Department of Electrical Engineering and Computer Science, University of Michigan,

1301 Beal Avenue, Ann Arbor, MI 48109-2122

Abstract

We show that the group delay in tunneling is not a traversal time but a lifetime of stored energy or stored probability escaping through both ends of the barrier. Because it is a lifetime associated with both forward (transmitted) and backward (reflected) fluxes, it cannot be used to define a group velocity for forward transit in cases where a wavepacket is mostly reflected. For photonic tunneling barriers the group delay is identical to the dwell time which is also a property of an entire wave function with reflected and transmitted components. Theoretical predictions and experimental reports of superluminal group velocities in barrier tunneling are re-interpreted.

winful@eecs.umich.edu

# I. INTRODUCTION

It is widely believed that particles or wavepackets can travel with superluminal velocity as they tunnel through a potential barrier [1-30]. The genesis of this belief can be traced to MacColl's 1932 paper which asserted, on the basis of an approximate solution to the time-dependent Schrödinger equation, that there is no appreciable delay in the transmission of the wavepacket through the barrier [1]. Later work by Bohm [31], Eisenbud and Wigner [32] provided a recipe for calculating the delay time in tunneling: namely that the energy derivative of the transmission phase shift yields the delay time, known variously as the phase time, Wigner time, or group delay. Hartman used this recipe to calculate the delay in tunneling through a rectangular barrier and found that the delay time for a tunneling particle is shorter than the delay time for a particle traversing the same distance in a barrier-free region [2]. Furthermore, in what has become known as the "Hartman effect", Hartman showed that the tunneling delay time saturates with increasing barrier length, at least until over-the-barrier energy components begin to dominate. A very natural (but, as it turns out, wrong) thing to do is to then take the length of the barrier $L$ and divide by the delay time $\tau_g$ to obtain a tunneling velocity $v_g = L/\tau_g$. The result of this exercise suggests that the group velocity in tunneling can become arbitrarily large as the barrier length increases while the delay stays fixed [5,10,14]. Such a result is not just an artifact arising from the use of the non-relativistic Schrödinger equation but is also present in the fully relativistic Dirac equation [19,26]. The presence of these rather large velocities occasions some discomfort as relativity forbids mass particles to travel faster than the speed of light in vacuum.

To test these predictions, a number of experiments were initiated in the early 1990's using electromagnetic waves which can tunnel as evanescent waves through forbidden regions in a manner analogous to that of quantum wavepackets [33,4,6,8,9,11]. The philosophy behind these tests is the identity between the time-independent Schrödinger equation for quantum particles and the Helmholtz equation for electromagnetic waves. These experiments, repeated by many [16-18,22,29], all show the result that the delay for the tunneling pulse is shorter than that of a non-tunneling pulse traversing the same distance. They also confirm that the group delay indeed describes the arrival of the peak of the tunneled pulse and that it saturates with barrier length. Indeed, even tunneling sound waves manifest similar "anomalous velocities" [34,35] (phenomena we might term supersonic), thus showing the generality of these phenomena as a consequence of wave behavior.

To rescue causality, superluminal tunneling velocities have been ascribed to a reshaping phenomenon [10,36-39]. The reshaping argument claims that the barrier transmits just the early part of the wavepacket and rejects the later parts, leading to a forward shift of the center of gravity of the pulse. The barrier, in effect, acts as a time-dependent shutter, resulting in a reshaped, narrowed, and "advanced" pulse. But, in fact, there is no experimental support for this reshaping argument. Experiments done with appropriately narrowband pulses show no pulse reshaping or pulse narrowing [18]. Yet, the reshaping argument appears to have gained currency and thus superluminal tunneling events are often casually dismissed as a mere reshaping phenomenon. No matter how these effects are produced, there is still the uncomfortable result that if group velocity describes the motion of a mass particle in quantum mechanics [31], then somehow the tunneling particle is able to travel much faster than it really ought to. This is not a tenable situation.

In recent papers we have suggested that the calculated and measured group delays in tunneling are not transit times [40-45]. Instead they are lifetimes of stored energy or integrated probability density leaking out of both ends of the barrier. In other words the barrier acts as an evanescent mode cavity with a finite lifetime. Because group delay describes an escape process through both channels, it should not be used as a measure of forward traversal time. In this paper we make these arguments more explicit and detailed. We demonstrate that *there is nothing superluminal, apparent or real, in barrier tunneling*. We show that the theoretical and measured delays are cavity lifetimes that describe the leakage of energy from both ends of the barrier. We prove the identity between group delay and cavity lifetime and support this proof with numerical simulations of energy decay within the barrier. Because the group delay refers to a bidirectional flux of energy, it cannot be assigned to a forward traversal velocity. Furthermore, there is no reshaping for narrowband pulses, which means that the "reshaping" mechanism that is widely believed does not really apply to these wavepackets.

Our focus is on wavepackets that tunnel without distortion, albeit with significant attenuation as a result of reflection. As we have shown [41,43], true tunneling is a quasi-static phenomenon requiring pulses whose spatial extent exceeds the length of the barrier. The quasi-static approximation says that the pulse temporal envelope changes very slowly in the time it takes a light front traveling at *c* to traverse the length of the barrier. As a result, at any instant in time, near steady-state conditions obtain.

## II. GROUP DELAY, FLUX DELAYS, AND DWELL TIME

A barrier acts as a filter, rejecting some frequency components and transmitting others. It also imparts a phase shift to each frequency component. The barrier is thus characterized by a complex transmission coefficient $|T(\omega)|e^{i\phi_t(\omega)}$ and a complex reflection coefficient $|R(\omega)|e^{i\phi_r(\omega)}$. These are steady state concepts that assume that the incident wave has existed long enough for the interferences that give rise to selective transmission to be established. Because the subject of tunneling time is still controversial [39,45], we will be purposely redundant in reviewing certain known properties of filters and time delay [46] as they pertain to tunneling. We will then introduce a new relation (Eqs.(13), (15)) between group delays, dwell time, and flux delays that demonstrates the status of group delay in tunneling as a measure of bidirectional energy transport. While the results apply equally to electromagnetic and quantum tunneling, we couch the description in terms of electromagnetic waves since the most convincing tunneling time experiments have been done with those waves.

If the Fourier transform of an incident time-dependent field $E_i(t)$ is $E_i(\omega)$, the transform of the transmitted field is

$$E_t(\omega) = E_i(\omega)|T(\omega)|\exp i\tilde{\phi}_t(\omega), \qquad (1)$$

where $\tilde{\phi}_t(\omega) = \phi_t(\omega) + k(\omega)L$ and $k$ is the wavenumber. Similarly the transform of the reflected field is

$$E_r(\omega) = E_i(\omega)|R(\omega)|\exp i\phi_r(\omega). \qquad (2)$$

Suppose the incident field is sufficiently narrowband that the magnitudes of the reflection and transmission coefficients are constant over the bandwidth of the pulse. Expanding the phase terms to first order about the center frequency $\omega_0$, we have

$$\phi_r(\omega) \approx \phi_r(\omega_0) + (\omega - \omega_0)\phi_r'(\omega_0) + O(\omega^2), \tag{3a}$$

$$\tilde{\phi}_t(\omega) \approx \tilde{\phi}_t(\omega_0) + (\omega - \omega_0)\tilde{\phi}_t'(\omega_0) + O(\omega^2). \tag{3b}$$

The transmitted field in the time domain is

$$E_t(t) = |T(\omega_0)| \exp i[\tilde{\phi}_t(\omega_0) - \omega_0 \tilde{\phi}_t'(\omega_0)] \int_{-\infty}^{\infty} E_i(\omega) \exp[-i(\omega[t - \tilde{\phi}_t'(\omega_0)])] d\omega$$

$$= |T(\omega_0)| \exp i[\tilde{\phi}_t(\omega_0) - \omega_0 \tilde{\phi}_t'(\omega_0)] E_i(t - \tau_{gt}), \tag{4}$$

where $\tau_{gt} = \tilde{\phi}_t'$ is the transmission group delay. Similarly, the reflected field is given by

$$E_r(t) = |R(\omega_0)| \exp i[\phi_r(\omega_0) - \omega_0 \phi_r'(\omega_0)] E_i(t - \tau_{gr}), \tag{5}$$

where $\tau_{gr} = \phi_r'$ is the reflection group delay. A filter with a constant amplitude transmission and a linear phase shift over the pulse bandwidth thus gives rise to a pure time delay without distortion. Any distortion or reshaping will have to come from the neglected higher order terms. Thus, "reshaping" cannot be seen as a mechanism for pure time delay. Since the power carried by the wave is $P(t) \propto |E(t)|^2$, for narrowband pulses the reflected and transmitted pulses are simply delayed and attenuated versions of the incident pulse:

$$P_t(t) = |T|^2 P_i(t - \tau_{gt}), \tag{6a}$$

$$P_r(t) = |R|^2 P_i(t - \tau_{gr}). \tag{6b}$$

The group delays can be related to power flow and stored energy in the barrier [42, 45]. For the electromagnetic barrier, Poynting's theorem provides a local statement of continuity of energy density and power flow:

$$\nabla \cdot \mathbf{S}(\mathbf{r},t) = -\frac{\partial u(\mathbf{r},t)}{\partial t}. \tag{7}$$

Here $\mathbf{S} = \mathbf{E} \times \mathbf{H}$ (W m$^{-2}$) is the Poynting vector and $u(\mathbf{r},t) = (\varepsilon \mathbf{E} \cdot \mathbf{E} + \mu \mathbf{H} \cdot \mathbf{H})/2$ is the electromagnetic energy density. For quantum wavepackets a similar continuity relation holds between the probability current $j(r,t)$ and the probability density $\rho(x,t)$:

$$\nabla \cdot \bar{j} = -\frac{\partial \rho}{\partial t}.$$

The integral form of Poynting's theorem then tells us that

$$-\oint_s \mathbf{S} \cdot \hat{\mathbf{n}} da = \frac{dU}{dt}, \tag{8}$$

with $U = \int_{vol} u dv$. Here, for the one-dimensional system the surface integral on the left hand side reduces to an integration over the input and output faces of the barrier. Thus evaluation of the surface integral in (8) yields

$$P_i(t) - P_r(t) - P_t(t) = \frac{dU}{dt}. \tag{9}$$

This relation simply says that the rate of increase of stored electromagnetic energy in the cavity is given by the incident power ($P_i$) minus the sum of the transmitted ($P_t$) and reflected ($P_r$) power. This equation represents a non-local relationship between incident and reflected fields on the one hand as measured at the input plane $z = 0$ and the transmitted field measured at the exit plane $z = L$, all measurements taken at the same instant $t$.

The group delay in transmission ($\tau_{gt}$) has the well defined meaning of the time at which a pulse peak appears at $z = L$, given that the peak of the incident pulse is at $z = 0$ at $t = 0$. Similarly, the group delay in reflection ($\tau_{gr}$) is the time at which a reflected pulse peak appears at $z = 0$, given that the peak of the incident pulse is at $z = 0$ at $t = 0$. Because of the absence of pulse distortion, the reflected and transmitted pulses are simply delayed, attenuated versions of the incident pulse, with possibly different delays. In practice, these pulses are measured some distance away from the barrier, but assuming free propagation outside the barrier region, the peaks can always be extrapolated to the boundaries of the barrier. To first order in the delay, assumed much smaller than the pulse width, we can write

$$P_i(t-\tau) \approx P_i(t) - \tau dP_i/dt .$$

Equation (9) then becomes

$$P_i(t) - |R|^2 \left[ P_i(t) - \tau_{gr} \frac{dP_i}{dt} \right] - |T|^2 \left[ P_i(t) - \tau_{gt} \frac{dP_i}{dt} \right] = \frac{dU}{dt} . \qquad (10)$$

Since $|R|^2 + |T|^2 = 1$ for a lossless barrier, Eq. (10) reduces to

$$\left[ |R|^2 \tau_{gr} + |T|^2 \tau_{gt} \right] \frac{dP_i}{dt} = \frac{dU}{dt} . \qquad (11)$$

Upon integrating both sides of Eq. (11) and setting the arbitrary constant equal to zero, we obtain the result

$$|R|^2 \tau_{gr} + |T|^2 \tau_{gt} = \tau_d \equiv \frac{U}{P_i} . \qquad (12)$$

The last identity above defines the dwell time $\tau_d$. The relation between the dwell time and the weighted sum of certain transmission and reflection delays is one that is often postulated as a criterion for the validity of those delays [47]. It has also been criticized as not being justified for quantum tunneling since it neglects interference terms which are significant at low particle energies [48]. For the photonic barrier in the slowly varying approximation, those interference terms disappear and thus (12) is an exact relation. For quantum tunneling the interference terms also disappear if the delays are averaged over the wavenumber spectrum of the incident wavepacket [47].

The dwell time is a weighted average of reflection and transmission delays and does not distinguish which of the two channels the energy stored from the input flux escapes through. Obviously, since it describes an escape process through *both* channels it cannot be ascribed to only the transmitted pulse nor used to assign a forward traversal velocity whenever there is some reflection. Upon dividing Eq. (12) by $\tau_d$, we obtain the relation

$$1 = \frac{\tau_{gr}}{\tau_r} + \frac{\tau_{gt}}{\tau_t}, \qquad (13)$$

where the transmitted flux delay is defined as

$$\tau_t = \frac{\tau_d}{|T|^2} = \frac{U}{P_t} \qquad (14a)$$

and the reflected flux delay is defined as

$$\tau_r = \frac{\tau_d}{|R|^2} = \frac{U}{P_r}. \qquad (14b)$$

The transmitted flux delay is the time it takes for all the stored energy in the barrier to leave the barrier in the direction of the transmitted flux. The reflected flux delay is the time it takes for all the stored energy to leave in the direction of the reflected flux. Equations (14) provide one way to distribute the dwell time between the reflection and transmission channels since it follows from the conservation law $|R|^2 + |T|^2 = 1$ that $1/\tau_r + 1/\tau_t = 1/\tau_d$ [49]. For a symmetric barrier it can be shown that $\tau_{gr} = \tau_{gt} \equiv \tau_g$ [47]. In that case Eq. (13) reduces to

$$\frac{1}{\tau_g} = \frac{1}{\tau_r} + \frac{1}{\tau_t}, \qquad (15)$$

and Eq. (12) becomes

$$\tau_g = \tau_d. \qquad (16)$$

Thus, as we have shown previously, the group delay and dwell time are identical for a symmetric photonic barrier. Equation (15) says that the sum of the rate of escape through the reflection channel and the transmission channel equals the inverse of the group delay. The group delay therefore represents the time it takes to empty the barrier through *both* channels.

### III. GROUP DELAY EQUALS CAVITY LIFETIME

An important point we wish to make is this: in the presence of reflections the group delay and the dwell time relate to the simultaneous escape of energy through both ends of the barrier. Neither of these times can be assigned to just the transmitted pulse or just the reflected pulse, in the sense of the time it takes a well defined pulse to travel from *A* to *B*. Indeed, the group delay is just the lifetime of stored energy escaping through both ends of the barrier. It is a cavity lifetime.

To see the connection to cavity lifetime, first recall the standard definition of the *Q* of a cavity [50]:

$$Q = \omega \frac{U}{P_d}, \qquad (17)$$

which is the time average stored energy divided by the power dissipated per cycle $P_d$. As a result of dissipation, the cavity mode has a finite lifetime, a 1/e lifetime of stored energy which is given by [50]

$$\tau_c \equiv \frac{Q}{\omega} = \frac{U}{P_d}. \qquad (18)$$

For a cavity with no internal losses the power dissipated is the power that escapes through the ends of the cavity. At steady state, this power lost equals the incident power, or $P_d = P_i$. Thus the cavity lifetime can be written

$$\tau_c = \frac{U}{P_i} = \tau_d, \qquad (19)$$

which shows that the cavity lifetime and the dwell time are identical. Furthermore, for a symmetric barrier all three quantities, the group delay, the dwell time, and the cavity lifetime are one and the same object.

As an important example, we consider a one-dimensional photonic bandgap structure with a refractive index profile $n(z) = n_0 + n_1 \cos(2n_0 \omega_B z/c)$, where $\omega_B$ is the Bragg frequency and $n_1$ is the amplitude of the index perturbation. For convenience we collect some known results [40]. The envelopes of the forward and backward fields propagating inside the structure satisfy the coupled-mode equations [51]

$$\frac{\partial E_F}{\partial z} + \frac{1}{v}\frac{\partial E_F}{\partial t} = i\kappa E_B e^{-i2\Delta\beta z}, \qquad (20a)$$

$$\frac{\partial E_B}{\partial z} - \frac{1}{v}\frac{\partial E_B}{\partial t} = -i\kappa E_F e^{i2\Delta\beta z}. \qquad (20b)$$

Here $\kappa = n_1 n_0 \omega_B / 2c$ is a coupling constant related to the strength of the refractive index perturbations, $\Delta\beta = n_0 \Omega / c$, $\Omega = \omega - \omega_B$, and $v = c/n_0$. The steady state solutions are

$$E_F(z) = E_0 [\gamma \cosh \gamma(z-L) + i(\Omega/v)\sinh \gamma(z-l)]/g, \qquad (21a)$$

$$E_B(z) = -i[E_0 \kappa \sinh \gamma(z-L)]/g, \qquad (21b)$$

where $\gamma = \sqrt{\kappa^2 - (\Omega/v)^2}$ and $g = \gamma \cosh \gamma L - i(\Omega/v)\sinh \gamma L$. The barrier amplitude transmission coefficient is $T = E_F(L)/E_o = (\gamma/|g|)e^{i\phi}$, the phase of which is given by

$$\phi_t = \tan^{-1}\left[(\Omega/\gamma v)\tanh \gamma L\right].$$

For an incident power

$$P_i = (1/2)\varepsilon_0 n_0 c |E_0|^2 A, \qquad (22)$$

the stored energy in the barrier is

$$U = U_0 \left[\frac{(\kappa^2/\gamma^2)(\tanh \gamma L)/\gamma L - (\Omega/\gamma v)^2 \operatorname{sech}^2 \gamma L}{1 + (\Omega/\gamma v)^2 \tanh^2 \gamma L}\right], \qquad (23)$$

where

$$U_0 = \frac{1}{2}\varepsilon_0 n_0^2 E_0^2 AL \qquad (24)$$

is the energy stored in a barrier-free region of the same length. From the expressions for stored energy and input power we calculate the group delay,

$$\tau_g = \frac{U}{P_i} = \tau_0 \left[ \frac{(\kappa^2/\gamma^2)(\tanh \gamma L)/\gamma L - (\Omega/\gamma v)^2 \mathrm{sech}^2 \gamma L}{1 + (\Omega/\gamma v)^2 \tanh^2 \gamma L} \right] = \frac{d\tilde{\phi}_t}{d\Omega}, \quad (25)$$

where $\tau_0 = L/v$. Figure 1 shows the normalized stored energy and group delay for a photonic bandgap structure as a function of detuning. The group delay is also the frequency-dependent cavity lifetime. Indeed at resonances of the barrier, where the transmission is unity, the expression for group delay reduces to

$$\tau_g = \frac{L}{v}\left[1 + \left(\frac{\kappa L}{m\pi}\right)^2\right], \quad (26)$$

which has the $L^3$ dependence first postulated for the cavity lifetime of distributed feedback lasers by Chinn [53]. Near these transmission resonances, constructive interference of the forward scattered wavelets results in an enhancement of the stored energy over the free-propagation value. Consequently, the group delay is greater than the free-propagation value in the vicinity of those resonances. Conversely, within the stop band, the interferences that give rise to coherent Bragg reflection result in a suppression of the stored energy below the value it would have had in the absence of the barrier. It is within this band that the group delay is less than the free-propagation value. It reaches a minimum value at midgap of

$$\tau_g = \frac{L}{v}\left[\frac{\tanh \kappa L}{\kappa L}\right]. \quad (27)$$

This delay can be arbitrarily short for strong enough barriers.

We will now show, by means of direct numerical solution of the coupled-mode equations, that the group delay is indeed the *1/e* lifetime of stored energy escaping through both ends of the barrier. The coupled-mode equations for forward and backward waves are integrated for a unit step function input until steady state is reached. The input is then switched off and the evolution of the integrated stored energy is monitored as a function of time. In Fig. 2 we show the decay of the stored energy for several values of the coupling strength $\kappa L$. Since the stored energy is numerically equal to the group delay for unit input power, we have normalized the plot of stored energy by the appropriate group delay. Hence what is plotted is $U(t)/\tau_g$. It is seen that the stored energy drops rapidly after turn off of the input. This rapid drop is largely due to the escape of the backward component through the input face of the barrier. Note that in steady state most of the energy is stored near the input end of the barrier (see Fig. 3b). The larger the coupling strength, the smaller the stored energy ($U \sim 1/\kappa L$) and thus the faster the decay. The temporal decay in the initial phase is approximately exponential. The thin horizontal line in the Figure indicates the *1/e* level of the stored energy. Upon reading off the values of time at the intersection of this line with the plots of *U(t)* we find that those times are close to the group delays calculated through Eq. (27). Thus the group delay is indeed the lifetime of stored energy leaving the cavity through both ends. After the initial rapid drop, there is a plateau which occurs because most of the energy in the backward wave has left, leaving behind the forward going component whose energy is approximately constant as it is no longer coupled to the backward wave. When the front of the forward component leaves the barrier at the exit after one transit time, the stored energy suddenly drops again. We have shown through the definition of Q and through direct numerical simulations that the group delay is the *1/e* lifetime of stored energy escaping through both ends of the barrier.

## IV. REINTERPRETATION OF TUNNELING EXPERIMENTS

With this interpretation of the group delay as a cavity lifetime, it is now possible to explain every aspect of "superluminal" tunneling experiments without ever mentioning the words "group velocity". In the typical pulsed tunneling time experiment, a pulse of electromagnetic energy is sent through a barrier-free region of length *L*. The arrival time of the peak of this pulse at a detector is used as a reference time. A barrier of length *L* is next inserted in the path of the pulse. The arrival time of the peak of a transmitted pulse is then compared with the reference time. Let the delay in traversing the barrier-free region be $\tau_0$. Let the delay due to the barrier be $\tau_1$. It is important to note that these experiments measure *time delays* (or, in interferometric measurements, mirror *positions* [6]). They do not directly measure velocity. Velocity is an inferred quantity. In order to relate a measured time delay to an inferred group velocity, one must ascertain that the measured delay is a propagation delay that can be assigned to the propagation of a forward pulse through a region.

Consider first the reference pulse. Figure 3a shows a snap shot of the energy density in the pulse as a function of position at the instant when the peak of the pulse arrives at the input plane. At that instant the flux of energy crossing that plane is the incident power $P_i$. Some time later, that flux of energy leaves the region. For the reference pulse, all the entering energy leaves the region at the exit plane *z = L* since there is no reflection or absorption. The time it takes for all the energy to leave is given by the stored energy divided by the rate at which energy enters: $\tau_0 = U_0 / P_i$. The time average stored energy in the transparent region of volume *V=LA* is just $U_0$ as given in Eq. (24). The net energy flux transmitted in the forward direction through this lossless, reflectionless region is just the input power $P_i$, given in Eq. (22). Upon dividing $U_0$ by $P_i$ we get $L/v \equiv \tau_0$, the time it takes for all the energy stored in the region of length $L$ to leave that region in the direction of the net flux and with velocity $v$. Here, because all the energy that enters leaves later in the forward direction, one can infer a sensible velocity $v = L/\tau_0$.

Now consider a pulse incident on a barrier. It should be noted right away that unlike the case of the reference pulse the incident and transmitted pulses are not the same entity. The incident pulse creates a cavity field which is made up of a sum of forward and backward propagating components that have undergone various amounts of multiple scattering within the barrier. This cavity field then gives rise to a transmitted pulse and a reflected pulse. The transmitted pulse is not the delayed incident pulse. It is the released barrier field. One cannot associate a given temporal point within the transmitted or reflected pulse with a given temporal point in the incident pulse. Again we calculate the stored energy in the barrier at the moment the pulse peak arrives. From Fig. 3b it is clear that this stored energy represented by the shaded area is much smaller than in the case of free space propagation. The energy density decays almost exponentially with distance. It consists of a forward and backward component whose sum, just inside the barrier exceeds the incident power. Most of the stored energy leaves the barrier in the backward direction and a small amount is transmitted in the forward direction. The group delay, the time it takes for this stored energy to escape through both forward and backward channels is $\tau_1 = U/P_i$. Because the stored energy is much smaller than in the free space case, for the same input power the delay time for this stored energy is much less than in the free space case. Note that this delay is not the time it takes for the input peak to propagate to the exit since the pulse does not really propagate through the barrier. What is really measured is the lifetime of stored energy escaping through both ends: the cavity lifetime. This explains why the group delay is shorter for stronger barriers (stronger barriers store less energy), why it saturates with barrier length (stored energy saturates with barrier length), and why it is less than the free space delay (barrier stores less energy than free space for the same input power). Because the group delay is associated with both forward and backward components it cannot be considered a delay time for forward traverse. Because it is not a delay time for forward transit, it is inappropriate to use this delay to define a group velocity for forward pulse propagation as distance *L* divided by delay. In short, $v_g \neq L/\tau_g$ when the delay is a delay associated with fluxes propagating in both directions. Even in regions of allowed propagation, when there are reflections and standing waves it becomes problematic to define a velocity of propagation and hence a traversal time. Only when the reflection is zero (as at transmission resonances) can one associate this delay with a forward traversal velocity.

We conclude this section with an analogy that may be useful. Imagine a vat filled with 100 liters of beer. If we pump out the beer at the rate 1 liter per second, how long will it take to empty the vat? 100 seconds, you say. Now suppose we only have 10 liters of beer in the vat and we still pump it out at 1 liter per second. This time it only takes 10 seconds to empty out. So, given the same pumping rate, the less beer in the vat, the less time it takes to empty it. Replace the vat with a photonic barrier, the beer with stored energy, and the pumping rate with power flow and it is easy to see why a pulse transmitted through a photonic barrier peaks sooner than one that passes through an equivalent length of free space.

## V. ABSENCE OF RESHAPING

The existence of "superluminal" group velocities in tunneling has been attributed to a reshaping phenomenon in which the barrier transmits the early parts of the incident pulse and rejects the later parts, acting in essence as a time dependent shutter [37-40]. This would imply that the transmission of the barrier is a function of time. A barrier whose transmission is time-dependent would necessarily distort an incident pulse. However, the exact numerical solutions of the coupled mode equations show that the transmission is the same for all parts of the delayed input pulse, at least over the detectable bulk of the pulse. This constancy of the transmission is also in agreement with experimental observations for pulses whose bandwidth is narrow compared to the stopband of the barrier.

Figure 4a shows the incident and transmitted pulses for a long pulse tunneling through a barrier of strength $\kappa L = 4$. In Fig. 4b, the two normalized pulses are overlaid so that their shapes can be compared. On this scale their shapes are identical. Notwithstanding claims to the contrary, there is no reshaping seen in theory or experiment. For a slowly varying input pulse, every portion of the main part of the transmitted pulse (after an initial transient) is delayed by the same amount from the incident pulse. Any distortion or reshaping is due to the higher order terms in the expansion of the transmission phase. It is not a mechanism for the prompt appearance of the transmitted peak. A pulse that is sufficiently long will not experience any reshaping. Thus "reshaping" cannot be seen as an essential part of tunneling dynamics. It is rather a sign of an approximation gone wrong. If the group delay is seen as a lifetime and not a transit time then there is no superluminality to explain away.

It does take time for the barrier reflectivity to build up to its steady state value. That build up process, however, occurs in the far wings of the input pulse, long before the main part of the pulse arrives. That part of the pulse, the front or "turn on" part contains the high frequency components which do not tunnel because they lie outside the stop band. That portion is necessarily "reshaped". However, that is not what is normally meant by the reshaping of a pulse. Furthermore, that portion has nothing to do with the tunneling process. Fig. 4c shows the evolution of the pulse front and the approach to quasi-steady state transmission. Note that the intensities here are 3-orders of magnitude smaller than in Figs. 4a and 4b which show the bulk of the pulse. The front propagates at $c$ and the reflectivity builds up in a couple of transit times.

## VI. IMPLICATIONS FOR QUANTUM TUNNELING

In the foregoing we have couched the interpretation of tunneling group delay in terms of electromagnetic wavepackets. Because of the analogy between the Schrödinger and Helmholtz equations, these results also hold for quantum wavepackets. Hence the group delay in quantum tunneling should also be seen as a lifetime and not a traversal time. The question then is, what is it a lifetime of? For electromagnetic waves for which the slowly varying envelope approximation holds, this lifetime was just the lifetime of stored energy escaping through both ends of the barrier. For quantum wavepackets the equivalent of stored energy in the barrier is the integral of the probability density in the barrier:

$$W = \int_0^L |\psi(x)|^2 \, dx,$$

where $\psi(x)$ is the wavefunction. For an incident particle flux $j_i$ the dwell time is given by [53,54]

$$\tau_d = W / j_i.$$

Mathematically the dwell time is the lifetime of the integrated probability density within the barrier irrespective of the escape channel. Since the integrated probability density is proportional to the number of particles, the dwell time can be also interpreted as the storage time of particles within the barrier, averaged over all incoming particles. If most of the particles are reflected, this lifetime will obviously be very short. For quantum wavepackets, in addition to the dwell time inside the barrier, there is a contribution to the tunneling delay time that arises from the interference between incident and reflected portions of the wavepacket in front of the barrier. The group delay takes into account this excess dwell time due to self-interference. The general relation between group delays and dwell times is now [42,45,55,56]

$$|R|^2 \tau_{gr} + |T|^2 \tau_{gt} = \tau_d + \tau_i,$$

where $\tau_i = -\operatorname{Im}(R)/kv$ is the self-interference delay. To emphasize its role as an additional dwell time we include the self-interference delay in an overall dwell time defined as

$$\tilde{\tau}_d = \tau_d + \tau_i.$$

For a symmetric barrier the group delay is thus seen as the duration of stored probability density (particles) in the barrier plus a contribution from any excess stored density due to interference in front of the barrier. Because the stored probability in front of the barrier arises from the interference between forward and backward waves, its contribution to the delay cannot be associated with either wave separately. The group delay is again a measure of bidirectional fluxes and cannot be associated with either the transmitted or reflected particles separately. At barrier resonances the group delay is just the lifetime of metastable states whereas for energies below the barrier height one can associate the delay with the lifetime of a highly transient virtual state.

For a wavepacket whose energy spread $\Delta E$ is much less than the barrier height $V_0$, the ratio of the limiting group delay to the wavepacket temporal extent $\tau_p$ is given roughly by

$$\tau_g / \tau_p = \delta x / \Delta x = \Delta E / V_0 \ll 1.$$

The spatial delay $\delta x$ caused by the barrier is a very small fraction of the uncertainty in the particle position $\Delta x$. Because this shift is such a small fraction of the wavepacket's length, a more meaningful measure of the duration of the tunneling process is the duration of the wavepacket itself.

## VII. CONCLUSIONS

For many years the anomalously short delays seen with tunneling quantum wavepackets and electromagnetic pulses have been taken to mean that these entities propagate with superluminal velocity through the barrier. We have now demonstrated that the delay time in barrier tunneling is actually the lifetime of stored energy (or stored particles) leaking through both ends of the barrier. Because it represents a bidirectional flow of energy it cannot be associated with a forward traversal time. Furthermore, we have shown that there is no pulse reshaping involved: every part of the main portion of the pulse suffers the same delay. This interpretation should help resolve some of the outstanding paradoxes in the physics of tunneling time.

**FIGURE CAPTIONS**

1. (Color online) Normalized group delay, dwell time, cavity lifetime, and stored energy versus frequency detuning for a symmetric photonic bandgap structure.

2. (Color online) Decay of the normalized stored energy versus time. The steady state incident beam is turned off at v$t/L$=20. Here $\Omega = 0$. The black dashed line shows the decay of stored energy in the barrier-free case.

3. (Color online) (a) Snapshot of a pulse traversing a region of free space. The shaded area shows the stored energy corresponding to the peak incident power.

   (b) Snapshot of a pulse interacting with a barrier. The shaded area shows the stored energy corresponding to the peak incident power. In front of the barrier is shown only the incident power but not the reflected power.

4. (Color online) (a) Incident, reference, and transmitted pulses for a barrier of strength $\kappa L = 4$.

   (b) Comparison of incident and transmitted pulse shapes.

   (c) Enlarged portion of Fig. 4b showing the evolution of the pulse front.

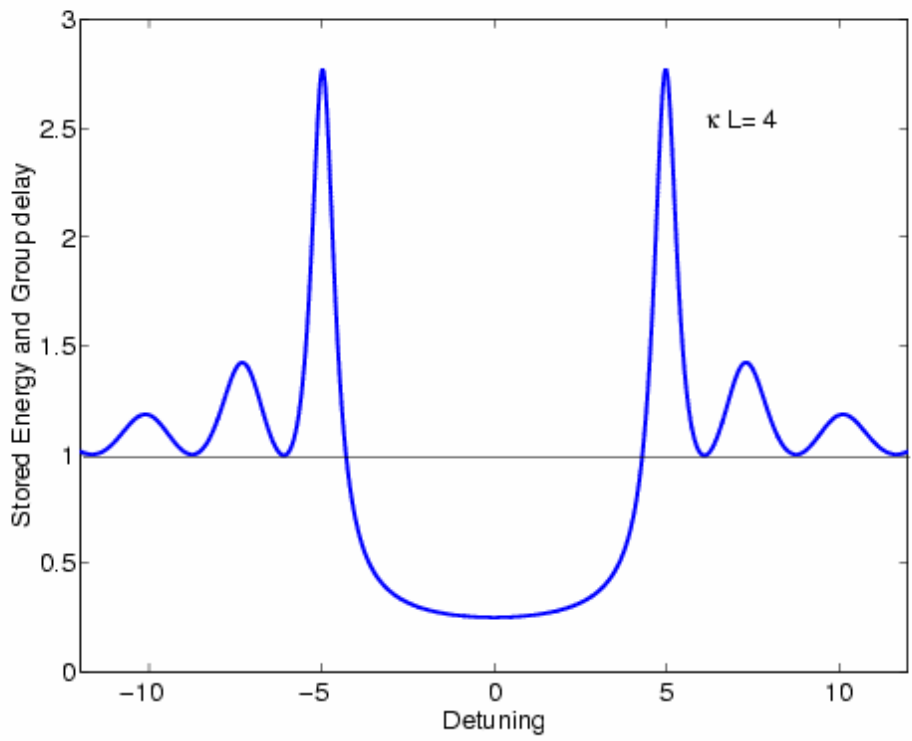

Winful Fig. 1

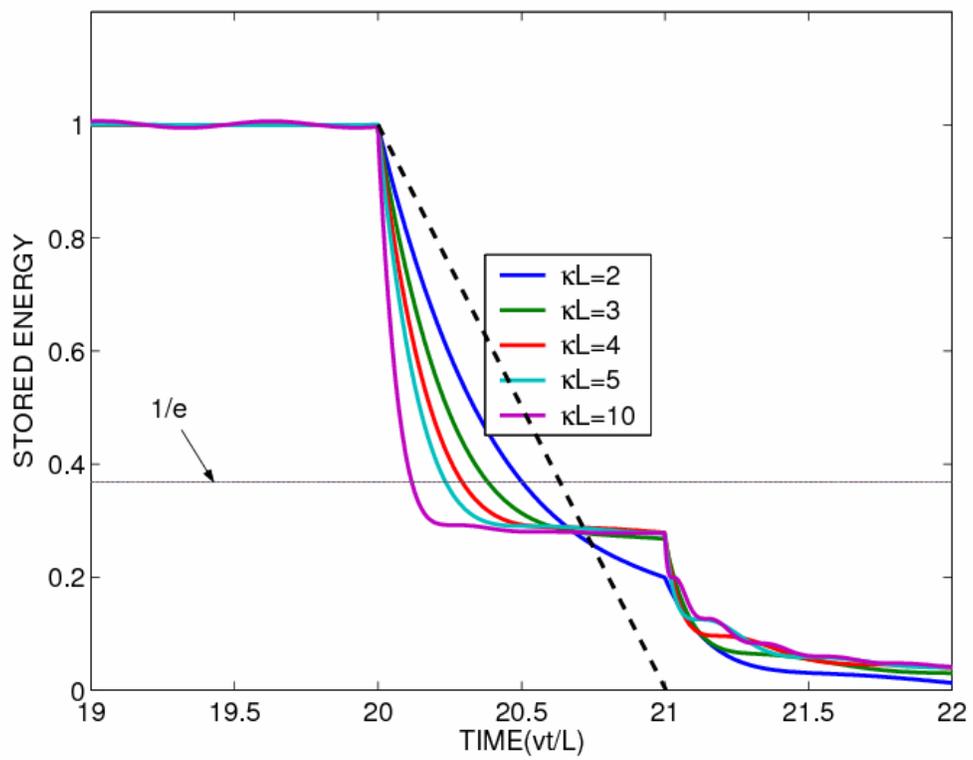

Winful Fig. 2

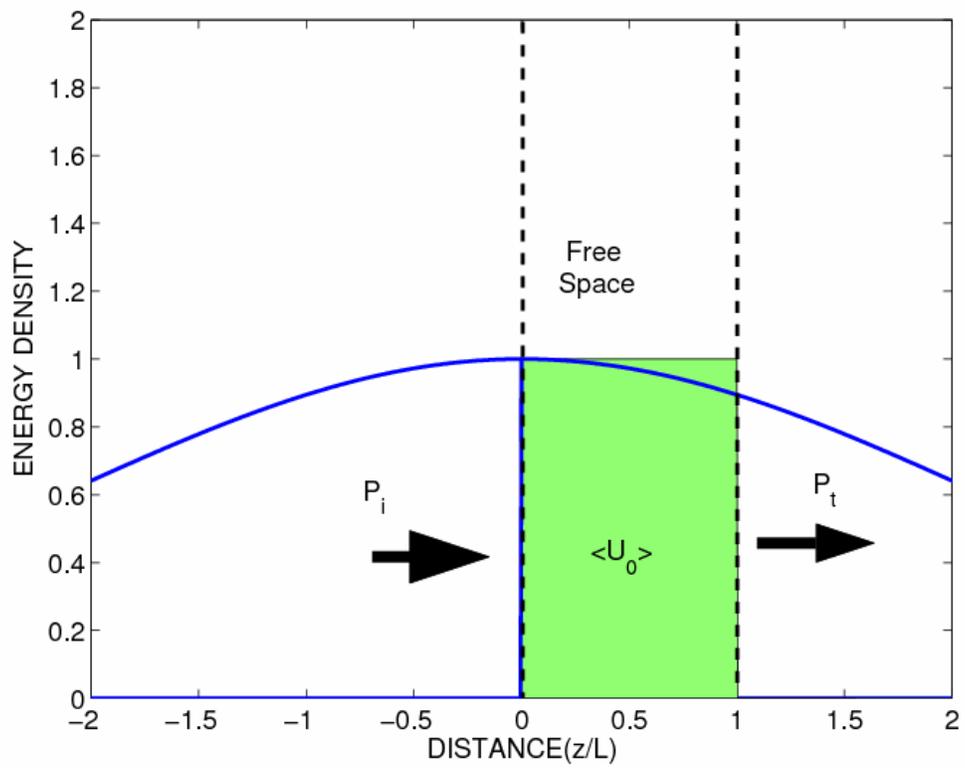

Winful Fig. 3a

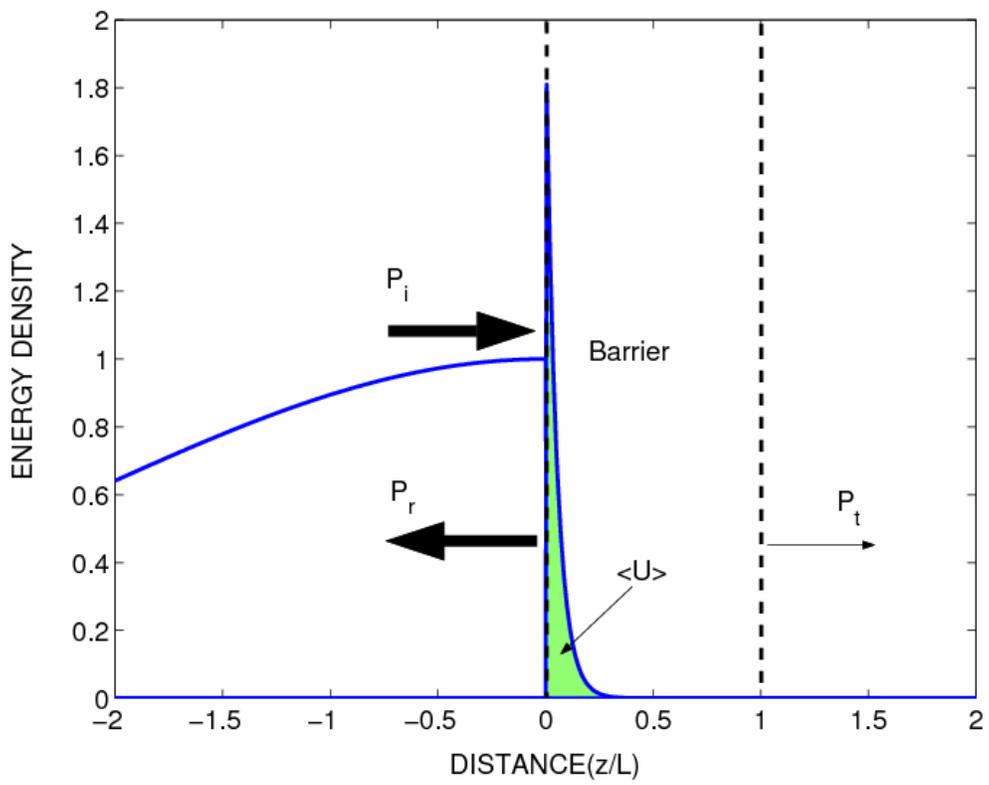

Winful Fig. 3b

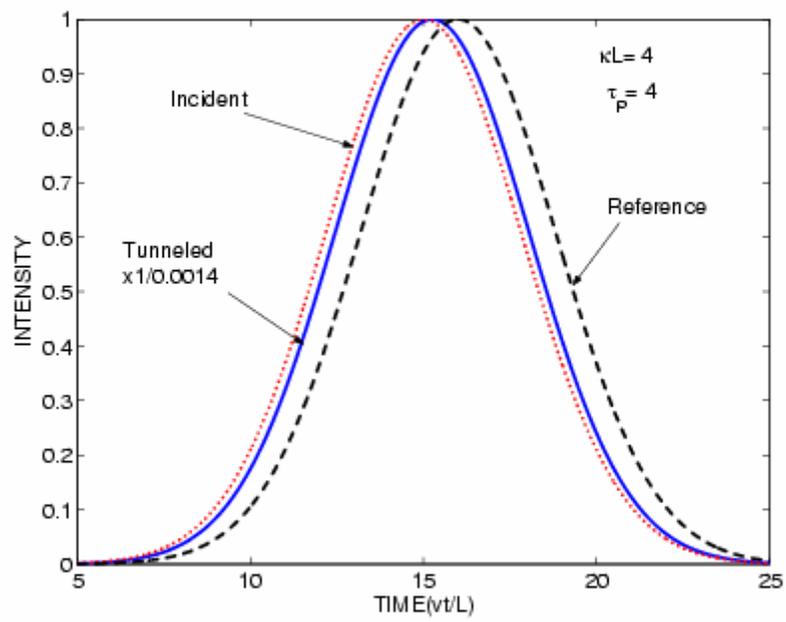

Winful Fig. 4a

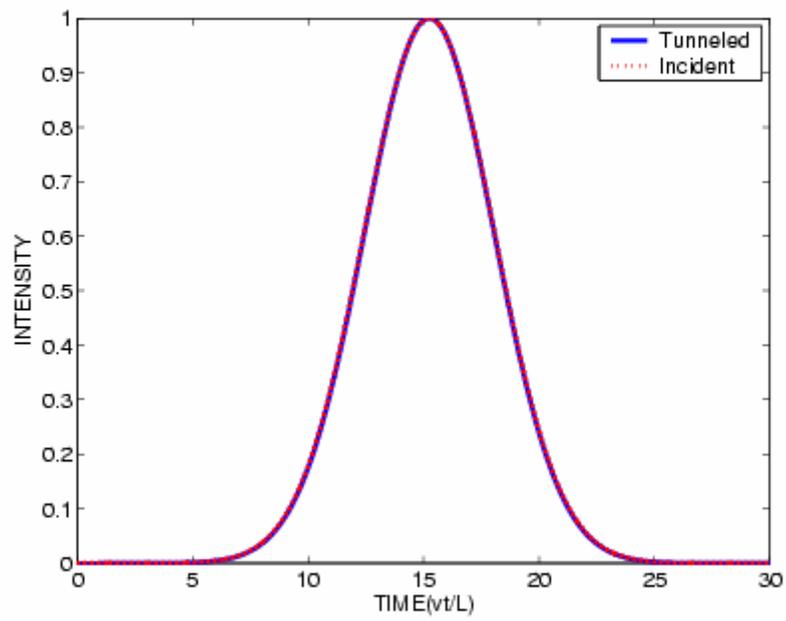

Winful Fig. 4b

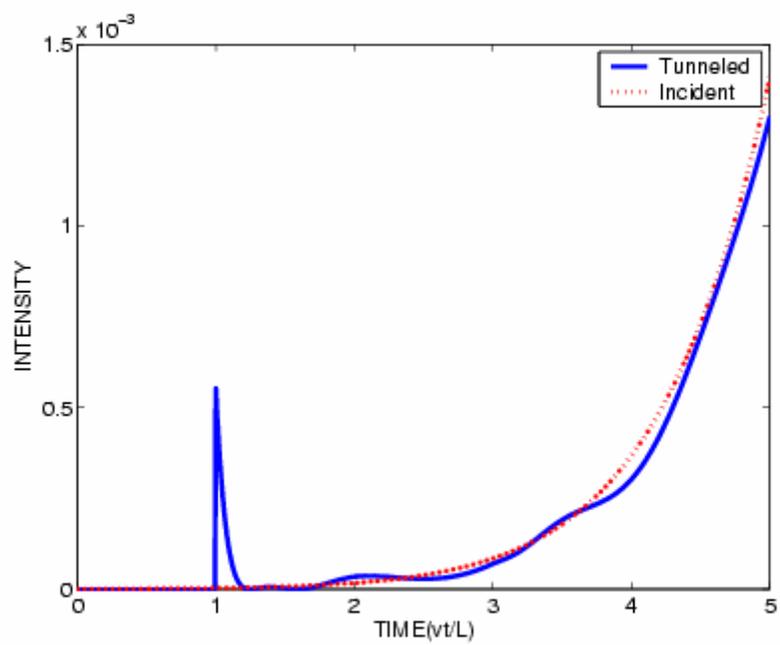

Winful Fig. 4c